\documentclass[%
showpacs,
 twocolumn, amsmath,amssymb,prl,aps
]{revtex4-1}

\usepackage{graphicx}
\usepackage{dcolumn}
\usepackage{bm}
\usepackage{subfigure} 

\def\bea{\begin{eqnarray}}
\def\eea{\end{eqnarray}}

\newcommand{\sgn}{\mathrm{sgn}}
\newcommand{\br}{{\bm r}}

\newcommand{\re}{\mathrm{e}}
\newcommand{\rd}{\mathrm{d}}
\newcommand{\bo}{\hat{b}^{\phantom\dag}}
\newcommand{\ba}{\hat{b}^{\dag}}

\newcommand{\no}{\hat{n}}
\newcommand{\Ho}{\hat{H}}
\newcommand{\Uo}{\hat{U}}
\newcommand{\la}{\langle}
\newcommand{\ra}{\rangle}
\newcommand{\be}{\begin{equation}}
\newcommand{\ee}{\end{equation}}
\newcommand{\bes}{\begin{eqnarray}}
\newcommand{\ees}{\end{eqnarray}}

\begin{document}


\title{Orbital-driven melting of a bosonic Mott insulator in a shaken optical lattice}

\author{Christoph Str\"ater}
\email{cstraeter@pks.mpg.de}
\author{Andr\'e Eckardt}
\email{eckardt@pks.mpg.de}
\affiliation{Max-Planck-Institut f\"ur Physik komplexer Systeme, N\"{o}thnitzer Stra{\ss}e 38, 01187 Dresden, Germany}

\date{\today}

\begin{abstract}
In order to study the interesting interplay between localized and dispersive orbital states in a system 
of strongly interacting ultracold atoms in an optical lattice, we investigate the possibility to 
coherently couple the lowest two Bloch bands by means of resonant periodic forcing. For bosons in one 
dimension we show that a strongly interacting Floquet system can be realized, where at every lattice 
site two (and only two) near-degenerate orbital states are relevant, whose tunneling matrix elements 
differ in sign and  magnitude. By smoothly tuning both states into resonance, the system is predicted to 
undergo an orbital-driven Mott-insulator-to-superfluid transition. As a consequence of kinetic 
frustration, this transition can be either continuous or first-order, depending on parameters such as 
lattice depth and filling. 
\end{abstract}

\pacs{37.10.Jk,03.75.Lm,67.85.-d,75.30.Mb}
\maketitle


\section{I. Introduction}

Orbital degrees of freedom play an important role in solid-state systems. A prominent example is
the intriguing physics of heavy-fermion compounds that emerges from the interplay between dispersive
conduction-band orbitals on the one hand and strongly localized orbitals, with a large effective mass 
and strong Coulomb interactions, on the other \cite{Hewson97,Coleman07,SiSteglich10,Gulacsi04}. However, 
in systems of ultracold atoms in optical lattices \cite{BlochDalibardZwerger08,LewensteinSanperaAhufinger}
orbital degrees of freedom, spanning Bloch bands above a large energy gap, are typically frozen out, at 
least in the interesting deep-lattice tight-binding regime where interactions are strong. Here, we 
investigate the possibility to coherently open on-site orbital degrees of freedom in a strongly 
interacting optical lattice system by means of near-resonant lattice shaking. We consider spinless 
bosons in one dimension (1D) and show how to realize a ``dressed-lattice'' system, where effectively at 
every lattice site the strongly localized ground-band orbital is nearly degenerate and coupled to the 
much more dispersive first-excited-band orbital. The tunneling matrix elements of the two orbitals 
differ strongly in magnitude and also in sign, with the latter leading to kinetic frustration. We 
predict an orbital-driven phase transition between a Mott insulator (MI) and a superfluid (SF) state 
when the population of the light orbitals is adiabatically increased by lowering the interorbital 
detuning. As a consequence of frustration and strong interorbital interactions, this transition is found 
to be either continuous or first-order, depending on parameters such as filling or lattice depth. 

In contrast to the present proposal, in previous experiments atoms were transferred non-adiabatically to 
excited bands of optical lattices by different methods \cite{GemelkeEtAl05,MuellerEtAl07,
SiasEtAl07,WirthEtAl11, OelschlaegerEtAl11,BakrEtAl11}. Moreover, lattice shaking has recently been 
employed for band-coupling in the weakly interacting regime, where condensation into two possible 
momentum states led to domain formation \cite{ParkerEtAl13}. Such band-coupling has been studied 
theoretically for non/weakly interacting particles and isolated sites \cite{ArlinghausHolthaus11,
ArlinghausHolthaus12,Sowinski12,ZhangZhou14,ZhengEtAl14,ChoudhuryMueller14,DiLibertoEtAl14}. Also 
orbtial coupling via magnetic resonances has been proposed \cite{PietraszewiczEtAl12} and there has been 
theoretical interest in the physics of excited orbitals not involving lower-lying states 
\cite{IsacssonGirvin05,WuEtAl06,Wu08a,Wu08b,LiEtAl12a,LiEtAl12b,PinheiroEtAl13}. Finally, the 
perturbative admixture of excited orbitals has been studied 
in theory \cite{LiEtAl06,LuehmannEtAl08,SchneiderEtAl09, Buechler10,HazzardMueller10, DuttaEtAl11,
LackiZakrzewski13, DuttaEtAl14} and experiment \cite{CampbellEtAl06,BestEtAl09,WillEtAl10,MarkEtAl11,
HeinzeEtAl11,JuergensenEtAl14}.


\section{II. Realizing the two-orbital model}
\begin{figure}[t]
\includegraphics[width=\linewidth]{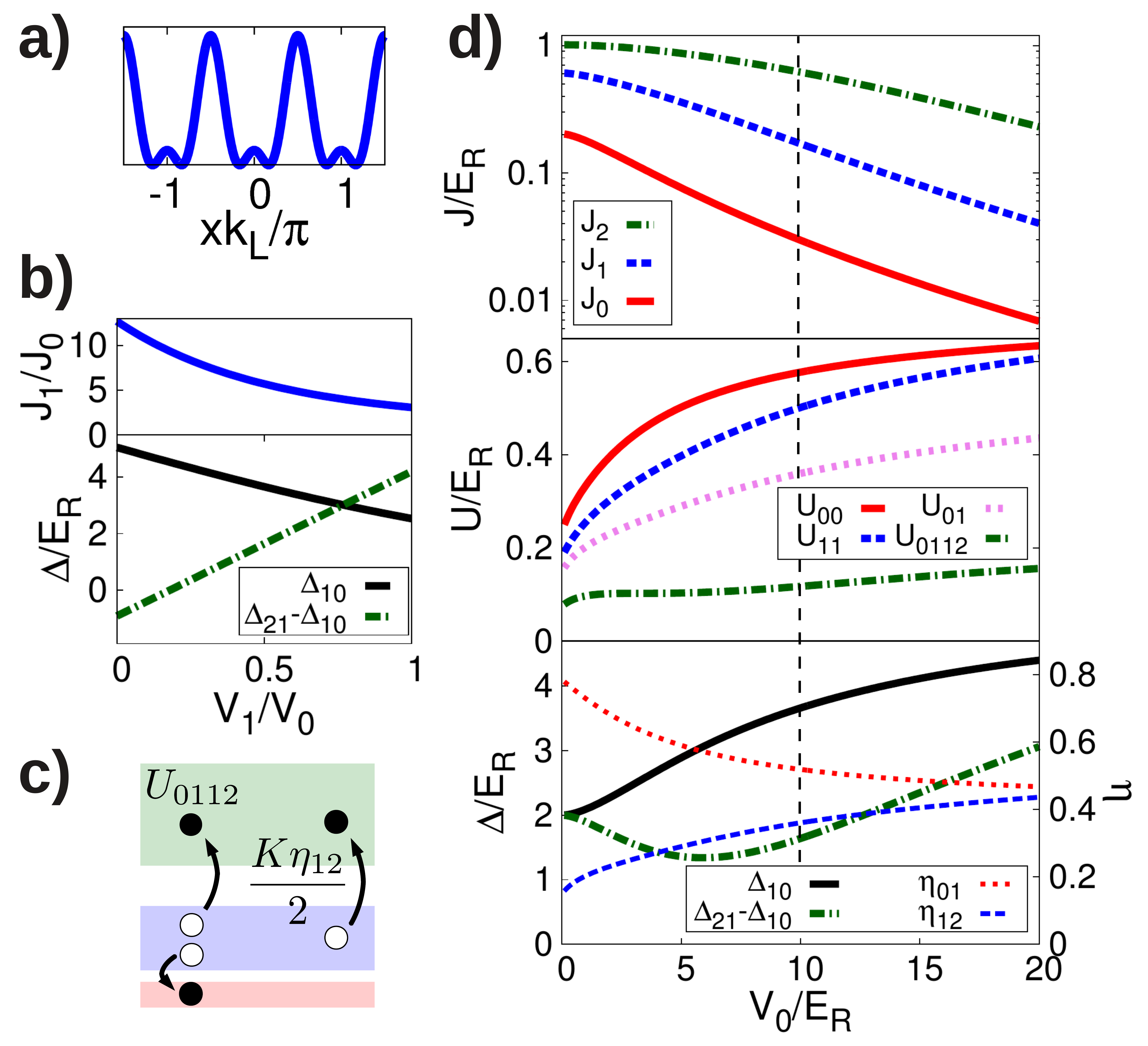}
\caption{(color online) 
(a) Dimerized lattice with $V_1/V_0=0.5$. 
(b) Impact of $V_1/V_0$ on $J_2/J_1$, $\Delta_{10}$, and $\Delta_{21}$, for $V_0/E_R=10$. 
(c) Most dominant loss channels, scattering into third band (zero-photon process) and
driving-induced coupling (single-photon process), are off-resonant due to dimerization. 
(d) Lattice parameters versus $V_0/E_R$ for $V_1/V_0=0.5$ and Rb$^{87}$.}
\label{fig:lattice}
\end{figure}

Consider spinless bosonic atoms of mass $m$ in an optical lattice 
$V(\br)=V_0\sin^2(k_L x)-V_1\sin^2(2k_Lx)+V_\perp[\sin^2(k_L y)+\sin^2(k_L z)]$. The $y$ and $z$ 
directions are frozen out by a deep lattice, we will assume $V_\perp=30E_R$, such that an array of
1D tubes with a dimerized lattice [Fig.~\ref{fig:lattice}(a)] is created. The recoil energy
$E_R=\hbar^2k_L^2/2m$ is needed to localize a particle on a lattice constant $\pi/k_L$. For Rb$^{87}$ a 
typical wave length of $2\pi/k_L= 852$ nm gives $E_R=2\pi\hbar\cdot 3.16$ kHz. 
Each 1D tube is described by the multi-band Bose-Hubbard Hamiltonian
$\Ho_0=\Ho_\text{kin}+\Ho_\text{os}$, where
\bes\label{eq:H0} 
\Ho_\text{kin} &=& 
-\sum_\ell\sum_\alpha (-1)^\alpha J_\alpha (\ba_{\alpha (\ell+1)}\bo_{\alpha \ell}  +\text{h.c.}),
\\
\Ho_\text{os}&=& 	\sum_\ell\bigg[\sum_\alpha \epsilon_\alpha \no_{\alpha \ell} 
			+\sum_{\{\alpha\}} \frac{U_{\{\alpha\}}}{2}
			\ba_{\alpha_1\ell}\ba_{\alpha_2 \ell}\bo_{\alpha_3\ell}\bo_{\alpha_4\ell}\bigg].
\ees
Here $\ba_{\alpha \ell}$ and $\no_{\alpha \ell}$ are the bosonic creation and number operator for a
Wannier orbital $w_\alpha(x-x_\ell)$ of Bloch band $\alpha=0,1,\cdots$, localized at
$x_\ell=\ell\pi/k_L$ 
\cite{*[{Since we consider a weak dimerization only, $V_1<V_0$, we do not need to employ 
generalized Wannier orbitals, localized in the left and right minimum of 
each double well [Fig.~\ref{fig:lattice}(a)], as in }] [] VaucherEtAl07}.
 The band-center energies and tunnel parameters fulfill
$0\equiv\epsilon_0<\epsilon_1<\cdots$ and $0<J_0<J_1<\cdots$, respectively. The interaction strengths 
$U_{\{\alpha\}}\equiv U_{\alpha_1 \alpha_2 \alpha_3 \alpha_4}
=(2\hbar^2a_sa_\perp^2/m)\int\!\rd x\, w_{\alpha_1}(x)w_{\alpha_2}(x)w_{\alpha_3}(x)w_{\alpha_4}(x)$
vanish for odd $\sum_i\alpha_i$, since $w_\alpha(x)=(-)^\alpha w_\alpha(-x)$, and depend on the
transverse localization length $a_\perp\simeq(V_\perp/E_R)^{1/4}/k_L$
and the scattering length $a_s$ ($\approx5.6$ nm for Rb$^{87}$). 

In the tight-binding regime, $\epsilon_1$ is typically much larger than the temperature and the chemical 
potential so that the orbital degree of freedom $\alpha$ is frozen out. We wish to coherently open this 
freedom by means of  time periodic forcing with near-resonant frequency $\hbar\omega\approx\epsilon_1$. 
In particular, the lowest band ($\alpha=0$) shall be coupled to the more dispersive first excited band
($\alpha=1$), without creating coupling to even higher-lying bands ($\alpha\ge2$). In order to achieve 
such a controlled situation---also in the regime where interactions are strong compared to tunneling---we 
combine two strategies: First we choose a driving scheme, namely sinusoidally shaking the lattice back and 
forth, that for weak forcing does not lead to multi-``photon'' interband transitions at resonances
$\Delta_{\alpha'\alpha}\equiv\epsilon_{\alpha'}-\epsilon_{\alpha}\approx m \hbar\omega$ with 
integer $|m|\ge2$. Second, we engineer the band structure by varying $V_1/V_0$ such that transitions to 
band 2 remain off resonant: With increasing $V_1/V_0$ the bands organize in pairs (0,1), (2,3), \ldots 
such that $\Delta_{10}$ and $\Delta_{32}$ as well as $J_1/J_0$ and $J_3/J_2$ decrease, while
$\Delta_{21}$ increases. For $V_0=10E_R$, already a slight dimerization $V_1/V_0=0.5$ ensures that 
$\Delta_{10}\approx3.7 E_R$ is noticeably smaller than $\Delta_{21}\approx 5.3 E_R$, rendering the
$\Delta_{21}$ transition off-resonant when $\hbar\omega \approx \Delta_{10}=\epsilon_1$. At the same 
time $V_1/V_0=0.5$ is small enough to keep a relatively large ratio $J_1/J_0\approx 5.7$, retaining the 
desired feature that $\alpha=0$ particles are much less dispersive than $\alpha=1$ 
particles [Fig.~\ref{fig:lattice}(b,d)].

By moving the lattice like $K k_L/(\pi m\omega^2)\cos(\omega t)$ in $x$ direction, an inertial 
force is created, described by
\be\label{eq:Hdr}
\Ho_\text{dr}(t) =K\cos(\omega t) \sum_\ell \bigg[\sum_{\alpha} \ell \no_{\alpha \ell} 
			+\sum_{\alpha'\alpha} \eta_{\alpha'\alpha}\ba_{\alpha'\ell}\bo_{\alpha \ell} \bigg].
\ee
Here $\eta_{\alpha'\alpha}=(k_L/\pi)\int\!\rd x\, w_{\alpha'}(x) x w_{\alpha_\alpha}(x)$ vanishes for 
even $\alpha'+\alpha$. We employ a time-periodic unitary transformation
$\Uo(t)=\exp(-i\sum_{\ell,\alpha}\no_{\alpha_\ell}\nu_\alpha\omega t)$ with 
integers $\nu_\alpha$, designed to shift all band energies $\varepsilon_\alpha$ to values
$\epsilon_\alpha'\equiv\epsilon_\alpha-\nu_\alpha\hbar\omega\in(-\hbar\omega/2, \hbar\omega/2]$ that are 
as close as possible to $\varepsilon_0=\varepsilon_0'=0$. This gives
$|\epsilon_1'|=|\epsilon_1-\hbar\omega|\ll\hbar\omega$ by choice of $\omega$ and
$|\epsilon_2'|=|\epsilon_2-2\hbar\omega|\sim\hbar\omega/2$ by choice of $V_1/V_0$; all other
$\epsilon_\alpha'$ are scattered somehow between $-\hbar\omega/2$ and $\hbar\omega/2$.
The periodic time dependence of the transformed Hamiltonian
$\Ho(t)=\Uo^\dag(\Ho_0+\Ho_\text{dr})\Uo-i\hbar\Uo^\dag\rd_t\Uo$ appears in the interband coupling 
parameters $K\eta_{\alpha'\alpha}\cos(\omega t) \re^{i(\nu_{\alpha'}-\nu_\alpha)\omega t}$ and
$U_{\{\alpha\}}\re^{i(\nu_{\alpha_1}+\nu_{\alpha_2}-\nu_{\alpha_3}-\nu_{\alpha_4})\omega t}$.
For weak forcing $K\ll\hbar\omega$ the driving frequency $\hbar\omega\sim\Delta_{10}$ is large compared
to the intraband terms as well as to the band coupling [Fig.~\ref{fig:lattice}(d)]. This allows to 
average the rapidly oscillating terms in the Hamiltonian over one driving period and to approximately 
describe the system by the effective time-independent Hamiltonian
$\Ho_\text{eff}=\frac{1}{T}\int_0^T\!\rd t\,\Ho(t)$, reading 
\bes\label{eq:Hmulti}
\Ho_\text{eff}&=& \Ho_\text{kin}
+\sum_\ell \bigg[\sum_{\alpha}\epsilon_\alpha'\no_{\alpha \ell} 
+\, K \sum_{\alpha'\alpha} \eta'_{\alpha'\alpha}
 \ba_{\alpha'\ell}\bo_{\alpha \ell} 
\nonumber\\&&
+\, \sum_{\{\alpha\}} \frac{U'_{\{\alpha\}} }{2}
\ba_{\alpha_1\ell}\ba_{\alpha_2 \ell}\bo_{\alpha_3\ell}\bo_{\alpha_4\ell}\bigg],
\ees
with $\eta'_{\alpha'\alpha}=\eta_{\alpha'\alpha}(\delta_{\nu_{\alpha'},\nu_{\alpha}+1}
+ \delta_{\nu_{\alpha'},\nu_{\alpha}-1})/2$ and $U'_{\{\alpha\}}=U_{\{\alpha\}} 
\delta_{\nu_{\alpha_1}+\nu_{\alpha_2},\nu_{\alpha_3}+\nu_{\alpha_4}} $. For a more systematic derivation 
of Eq.~(\ref{eq:Hmulti}), $\Ho_\text{eff}$ is defined as the generator of the time evolution over one 
period \cite{Shirley65} and computed using degenerate perturbation theory in the extended Floquet 
Hilbert space \cite{Sambe73}, similar like in Refs.~\cite{EckardtHolthaus07,EckardtHolthaus08b}. In 
leading order one recovers Eq.~(\ref{eq:Hmulti}). The leading correction contains tiny second-order 
coupling to bands $\alpha\ge3$ of order $c^2/\hbar\omega$ to be neglected, where $c\lesssim 0.1E_R$ is 
a  typical interband coupling matrix element and $\hbar\omega\gtrsim3E_R$. 

It is a crucial property of lattice shaking (\ref{eq:Hdr}) that in $\Ho_\text{eff}$ the interband 
coupling $K\eta_{\alpha'\alpha}'$ is a single-photon process, with $\nu_{\alpha'}=\nu_\alpha\pm1$, and 
that scattering $U_{\{\alpha\}}'$ is a zero-photon process, with
$\nu_{\alpha_1}+\nu_{\alpha_2} = \nu_{\alpha_3}+\nu_{\alpha_4}$. No multi-photon processes are found for weak 
driving. Thus, in $\Ho_\text{eff}$ above the bands 0 and 1 are coupled to band 2 only, 
via the processes sketched in Fig.~\ref{fig:lattice}(c). These processes are, however, off-resonant, since
$\epsilon_2'\sim\hbar\omega$. The bands 0 and 1 are, therefore, to good approximation isolated and 
described by the two-band (2B) model
\bes\label{eq:Heff}
\Ho_{2B} &=&\Ho_\text{kin} +\sum_{\ell}\Big[\delta \no_{1\ell} 
	-\gamma(\ba_{1\ell}\bo_{0\ell}+\text{h.c.})+ 2U_{10}\no_{0\ell}\no_{1\ell}
\nonumber\\&&
+\,\frac{U_{00}}{2}\no_{0\ell}(\no_{0\ell}-1)+\frac{U_{11}}{2}\no_{1\ell}(\no_{1\ell}-1)
  \Big],
\ees
where $\gamma=-K\eta_{10}/2$, $U_{\alpha'\alpha}\equiv U_{\alpha\alpha'\alpha'\alpha}$ and
$\delta=\Delta_{10}-\omega$. For $V_0=10E_R$, $V_1/V_0=1/2$, and $K=0.5E_R$ we obtain
$J_0\approx0.030E_R$, $J_1\approx0.17E_R$, $\gamma\approx0.13E_R$, $U_{00}\approx0.58E_R$, 
$U_{01}\approx0.36E_R$, and $U_{11}\approx0.50E_R$.

\begin{figure}[t]
\includegraphics[width=\linewidth]{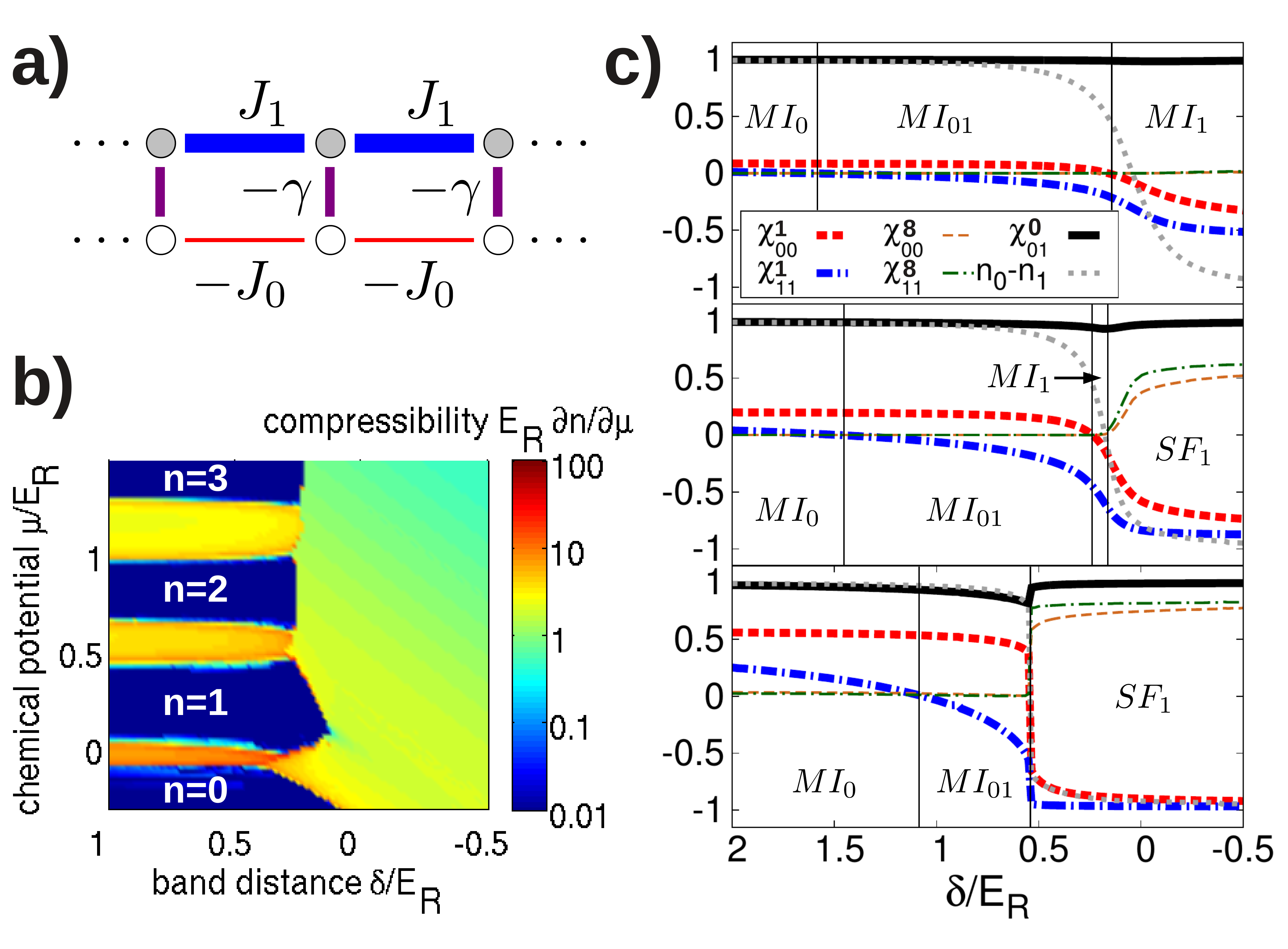}
\caption{ (color online)
(a) Sketch of the effective model, grey (white) circles correspond to the $\alpha = 1$ (0) state at each 
site $\ell$, with energy $\delta$ (0). 
(b) Ground-state compressibility $\partial_\mu n E_R$ in the $\mu$-$\delta$ plane for $V_0/E_R=10$ and
$V_1/V_0=1/2$. 
(c) Correlations $\chi^{\ell}_{\alpha'\alpha}\equiv \la\ba_{\alpha'\ell}\hat{b}_{\alpha 0} \ra /
\sqrt{n_{\alpha'}n_\alpha}$ and imbalance $n_0-n_1$ for fixed filling $n=1$ versus $\delta$, 
with $V_1/V_0 =0.5$ and $V_0/E_R=15, 10, 5$ (from top to bottom). Numerical data in (b) and (c) obtained 
for $M=30$ rungs under periodic boundary conditions using TEBD in imaginary time \cite{TEBD,Vidal04}, 
with bond dimensions 14 (b) and 30 (c).} 
\label{fig:phases}
\end{figure}

$\Ho_{2B}$ describes a highly tunable 1D ladder system [Fig.~\ref{fig:phases}(a)] with interesting
properties: The tunneling matrix elements along both legs 
(i.e.\ in both bands) differ in sign and magnitude. The former leads to maximal kinetic frustration with 
a flux of $\pi$ per plaquette \cite{Honerkamp03,HottaFurukawa06,EckardtEtAl10,StruckEtAl11,GrescherEtAl13,
TielemanEtAl13,DharEtAl13,YudinEtAl14}. The latter
makes leg 0 more prone to localization than leg 1. The hybridization of both legs is controlled by the 
energy separation $\delta$ and the coupling $\gamma$, which can be tuned via the frequency and strength 
of the driving, respectively. Finally, the system features strong interorbital interactions $U_{10}$, 
with two-particle energies $2U_{01}>U_{00}>U_{11}$. 

In order to investigate the 2B model (\ref{eq:Heff}), the following experimental protocol can be 
pursued. After the system is prepared in (or close to) the undriven ground state, populating band 0, the 
driving strength $K$ is ramped up smoothly to the desired value. During this step the detuning 
$\delta$ is still large enough to suppress any significant occupation of band 1. Then, the orbital 
freedom is opened by smoothly lowering $\delta$.

\section{III. Orbital-driven Mott transition}

We study the ground state of $\Ho_{2B}$ versus $\delta$. For large positive
(negative) $\delta/|\gamma|$ only leg 0 (leg 1) will be occupied; the system effectively reduces to a 
1D Bose-Hubbard chain. For integer filling of $n$ particles per site, the ground state of such a chain 
with tunneling $J$ and on-site repulsion $U$ is a gapped (i.e.\ incompressible) MI with localized 
particles if $J/U<(J/U)^{(n)}_c$, where $(J/U)^{(1)}_c\approx0.26$ \cite{ElstnerMonien99}. Otherwise, it 
is a gapless SF with quasilong-range order. Thus, for $n=1$ the system is a MI for
$\delta\gg|\gamma|$, since $J_0/U_{00}\approx 0. 051$, and a SF for $-\delta\gg|\gamma|$, since
$J_1/U_{11}\approx 0.34$.

It is instructive to control $n$ via the chemical potential $\mu$, introduced by adding  
$-\mu\sum_\ell(\no_{0\ell}+\no_{1\ell})$ to $\Ho_{2B}$. In Fig.~\ref{fig:phases}(b), we plot the
ground-state compressibility $\partial_\mu n$ in the $\mu$-$\delta$ plane, computed by time-evolving 
block decimation (TEBD) in imaginary time \cite{TEBD,Vidal04}. As expected \cite{ElstnerMonien99}, for
$\delta/|\gamma|\gg 1$ we find incompressible MI phases at integer filling $n$, interrupted by SF phases 
where $n$ changes in $\mu$ 
direction, while for $-\delta/|\gamma|\gg 1$ the system is a compressible SF. 
When $\delta$ is lowered, the filling $n_0$ ($n_1$) of leg 0 (1) decreases (increases). In response,
an orbital-driven transition occurs, either between a MI and a SF or, for fractional filling, between 
different SFs. For the given parameters, these are first-order transitions, except at the tip of the $n=1$
Mott phase, where a continuous transition is found. The discontinuous SF-to-SF transition, where the 
ground state changes abruptly, happens when near $\delta=2(J_0-J_1)\approx-0.28 E_R$ a boson suddenly 
prefers to delocalize with quasimomentum $\pi$ in leg 1, rather than with quasimomentum 0 in leg 0. The 
discontinuous MI-to-SF transition, to be explained below, is more subtle. 

A strong-coupling argument explains the orbital-driven MI-to-SF transition. Within the MI state, $n_1$ 
increases smoothly when $\delta$ is lowered, and the larger $n_1$ the larger is the reduction of kinetic 
energy $\approx 2J_1(n_1+1)$ (or $\approx 2J_1n_1$) that a particle (or a hole) acquires by delocalizing 
along leg 1 on the MI background. When the kinetic energy reduction of a particle-hole excitation exceed 
its interaction-energy cost $\approx (U_{11}+2U_{01}n_0\delta_{n,1})$, these excitations proliferate and 
the ground state becomes a SF as seen in Fig.~\ref{fig:phases}(b). This transition can also be observed 
in Fig.~\ref{fig:phases}(c,middle) where we plot $n_0-n_1$ and the ground-state correlations
$\chi^{\ell}_{\alpha'\alpha}\equiv\la\ba_{\alpha'\ell}\bo_{\alpha0}\ra/\sqrt{n_{\alpha'}n_\alpha}$ 
versus $\delta$, for the same parameters and sharp filling $n=1$ 
\footnote{Experimentally $n_0$ and $n_1$ and the Fourier transforms of $n_0\chi^{\ell}_{00}$ and
$n_1\chi^{\ell}_{11}$ can be measured via band mapping \cite{MuellerEtAl07}.}.
While $|\chi_{11}^{\ell}|$ decays exponentially with $\ell$ in the MI phase, in the SF regime the decay 
is only algebraic. Therefore, the transition is indicated by a significant increase of correlations on 
longer distances such as $\chi^8_{\alpha\alpha}$. It is found near $\delta= 0.2E_R$, in fair agreement 
with the above estimate giving $n_0-n_1\approx(8J_1-U_{11}-2U_{01})/(4J_1+2U_{01})\approx 0.10$.

In Fig.~\ref{fig:phases}(c, middle) we can identify three different types of MI states, characterized by 
respective signs $s_\alpha\equiv\sgn(\chi^1_{\alpha\alpha})$ of the short-range correlations along both 
legs. This is a consequence of kinetic frustration; while the tunneling matrix elements $-J_0$ and $J_1$ 
favor $s_0=+1$ and $s_1=-1$, the rung coupling $-\gamma$ favors $s_0=s_1$. We use the label MI$_{01}$ if 
both legs retain their favored correlations ($s_0=-s_1=+1$), and MI$_\alpha$ if leg $\alpha$ 
dominates the other one [$s_0=s_1=+1$ (-1) for $\alpha = 0$ (1)]; similar labels are used for SF 
states. Due to the strong interleg interactions $2U_{01}>U_{00},U_{11}$ the system does not feature the 
chiral time-reversal symmetry broken MI or SF ground states with complex $\chi^\ell_{\alpha'\alpha}$ 
predicted in Ref.~\cite{DharEtAl13}. Treating both $\gamma$ and the $J_\alpha$ as perturbation the
MI$_0$-to-MI$_{01}$ transition is predicted to occur at 
$\delta=U_{00} J_1/(2J_0) \approx1.6$ \cite{Supplemental}.
Experimentally, this transition is hardly observable, since it occurs at tiny
$n_1\simeq|\gamma/\delta|^2\approx 0.007$. The MI$_{01}$-to-MI$_1$, happening when 
$\delta$ is lowered further, is of greater importance. A perturbative treatment of the tunneling matrix 
elements $J_\alpha$, neglecting $\gamma$ and $\delta$ on interaction-dominated doubly occupied sites, 
predicts this transition to occur when
$n_0-n_1\approx[(J_1-J_0)U_{00}-4J_0U_{01}]/[(J_1-J_0)U_{00}+4J_0U_{01}]\approx0.31$ \cite{Supplemental},
in reasonable agreement with the numerics. 
These transitions can be observed also in a deeper lattice, where the system remains a MI for small
$\delta$ [Fig.~\ref{fig:phases}(c,top)].

For a lower lattice depth of $V_0/E_R=5$, the MI-to-SF transition occurs earlier and already within the
MI$_{01}$ regime [Fig.~\ref{fig:phases}(c,bottom)]. This is explained by the above estimates that 
predict the MI-to-SF transition to occur when $n_0-n_1\approx 0.94$, well before the estimated value
$n_0-n_1\approx 0.23$ for the MI$_{01}$-to-MI$_1$ transition is reached. As a consequence, the MI-to-SF 
transition is rendered discontinuous. The discontinuity results from an abrupt change in the structure 
of the short-range correlations along leg 0. Namely, the SF phase is of SF$_1$ type, with 
$s_0=s_1=-1$, such that the short-range correlations along leg 0 have to undergo a finite jump at the
MI$_{01}$-to-SF$_1$ transition. 
The same argument also explains the first-order nature of the orbital-driven MI-to-SF transitions for
$V_0/E_R=10$ at filling $n\ge2$, visible as a sharp jump of the compressibility in
Fig.~\ref{fig:phases}(b). 
All in all, the fact that the orbital-driven MI-to-SF transition can be discontinuous results from the 
combination of kinetic frustration, tunneling imbalance $J_1\gg J_0$, and strong interband interactions
$U_{01}$, all stemming from the spatial structure of the Wannier orbitals.

\section{IV. Preparation dynamics and heating}

\begin{figure}[t]
\includegraphics[width=\linewidth]{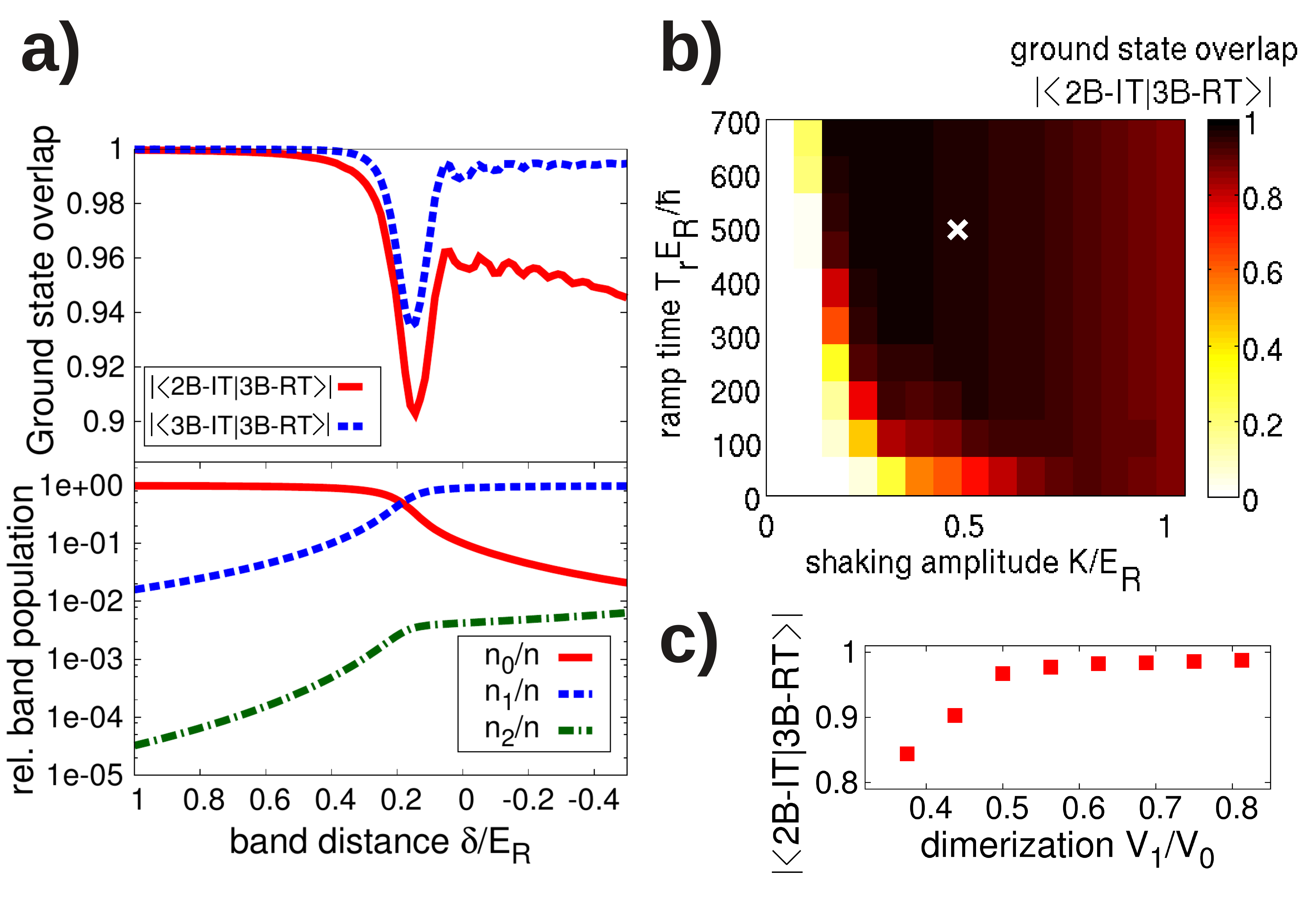}
\caption{ (color online)
Real-time (RT) evolution of the effective model (\ref{eq:Hmulti}) with three bands (3B) $\alpha=0,1,2$,  obtained by TEBD \cite{TEBD,Vidal04} with bond dimensions 24 (a), 22 (b), and 22-40 (c). 
(a) Occupations and overlaps with the [imaginary time-evolved (IT)] ground states of the 3B and 
2B model. Starting in the ground state at $\delta=\epsilon_1-\omega=1E_R$, $\delta$ is lowered 
linearly to $\delta=-0.5E_R$ within a time of $T_r=500 \hbar/E_R$; for $n=1$, $V_0/E_R=10$, $V_1/V_0=0.5$,
$K/E_R=0.5$ and $M=16$ rungs under periodic boundary condtions. 
(b) After-ramp overlap for $M=10$, like in (a) but varying $T_r$ and $K$. 
(c) After-ramp overlap for $M=10$, like in (a), but varying $V_1/V_0$.}
\label{fig:prep}
\end{figure}

When $\delta$ is lowered slow enough, the system is expected to approximately follow the ground state 
of the 2B model (\ref{eq:Heff}), unless the first-order transition is crossed. This desired dynamics is 
effectively adiabatic \cite{EckardtHolthaus08b}, i.e.\ adiabatic with respect to $\Ho_{2B}$, but 
diabatic with respect to tiny coupling matrix elements neglected in $\Ho_{2B}$. We have simulated the 
time evolution of the system [Fig.~\ref{fig:prep}(a)] using TEBD \cite{TEBD,Vidal04}. For parameters 
like in Fig.~\ref{fig:phases}(c,middle), $\delta/E_R$ is ramped from $1$ to $-0.5$ within a time
$T_r=500\hbar/E_R\approx 25$ms. In order to probe the validity of the 2B model (\ref{eq:Heff}), we 
include the major ``loss'' processes depicted in Fig.~\ref{fig:lattice}(c) by employing Hamiltonian
(\ref{eq:Hmulti}) with three bands ($\alpha=$0,1,2). In Fig.~\ref{fig:prep}(a), one can clearly see that 
the $\alpha=2$ occupation $n_2$ remains very low and that the overlap with the 
instantaneous 2B ground state stays close to 1. Both clearly shows that the driving does not cause 
detrimental heating and justifies a description of the driven system in terms of the 2B model
(\ref{eq:Heff}). It, moreover, indicates that an effectively adiabatic time evolution is possible, 
despite a noticeable dip of the overlap at the Mott transition (resembling the behavior of Landau-Zener 
sweeps \cite{LimBerry91}). Thus, the protocol allows for the preparation of stable low-entropy states in 
an excited Bloch band. 

The overlap plotted in Fig.~\ref{fig:prep}(b) versus $T_r$ and $K$ measures the effective adiabaticity. 
Too small $T_r$ and $\gamma\propto K$ spoil the adiabatic dynamics within the 2B model and for too large 
$K$ the coupling to band 2 becomes relevant. Moreover, for too large $K$ and $T_r$ 
slow second-order loss processes (not included) can occur. Finally, Fig.~\ref{fig:prep}(c) shows that 
for strong interactions a minimal dimerization of $V_1/V_0=0.5$ is crucial. Different from the weakly 
interacting case \cite{ParkerEtAl13}, we find significant transfer to the second excited band 2 for the 
simple cosine lattice ($V_1/V_0=0$).

\section{V. Conclusion and outlook}
We have shown that lattice shaking is a feasible tool to coherently open on-site orbital degrees of 
freedom in a strongly interacting optical lattice system and that the interplay between Wannier orbits 
of different structure gives rise to rich physics already for spinless bosons in 1D.
Extending the scheme to spinful fermions, the interplay between strongly localized and dispersive orbital 
states should permit to mimic aspects of the intriguing heavy-fermion physics and to realize
periodic-Anderson-like models \cite{Hewson97,Coleman07,SiSteglich10,Gulacsi04}. The extension to higher 
dimensional lattices should provide a feasible scheme for the preparation of low-entropy states in 
excited bands as they have been discussed before \cite{IsacssonGirvin05,WuEtAl06,Wu08a,Wu08b,LiEtAl12a,
LiEtAl12b,PinheiroEtAl13} and, moreover, to couple them to strongly localized lowest-band orbits. 
Finally, by employing sufficiently off resonant forcing, keeping $\delta$ large enough, one might 
enhance and control the perturbative admixtures of excited bands \cite{LiEtAl06,LuehmannEtAl08,
SchneiderEtAl09, Buechler10,HazzardMueller10, DuttaEtAl11, LackiZakrzewski13, DuttaEtAl14,CampbellEtAl06,
BestEtAl09,WillEtAl10,MarkEtAl11,HeinzeEtAl11,JuergensenEtAl14}, e.g.\ in order to enhance superexchange 
processes by engineering density-dependent tunneling.

\begin{acknowledgments}

We thank Mikl{\'o}s Gul{\'a}csi for discussion. CS is grateful for support by the Studienstiftung des 
deutschen Volkes.

\end{acknowledgments}


\section{Appendix A: Transition between MI$_0$ and MI$_{01}$}
For large negative $\delta$, one can treat both $\gamma$ and the tunneling matrix elements $J_\alpha$ as 
a perturbation. For unit filling $n=1$ the unperturbed ground state takes the simple form
\be
|\psi_0\ra = \prod_\ell \ba_{0\ell}|\text{vac}\ra
\ee
with the vacuum state $|\text{vac}\ra$. A finite correlation $\la\ba_{10}\ba_{11}\ra$ will appear in 
third order. Namely one has
\bes
\la\psi|\ba_{10}\bo_{11}|\psi\ra 
&\simeq& \la\psi_1|\ba_{10}\bo_{11}|\psi_2\ra+\la\psi_2|\ba_{10}\bo_{11}|\psi_1\ra
\nonumber\\
&=&2\la\psi_1|\ba_{10}\bo_{11}|\psi_2\ra
\ees
with $|\psi_k\ra$ denoting the state correction appearing in $k$th order perturbation theory. 
Here the relevant term of $|\psi_1\ra$ reads
\be
\frac{-\gamma \ba_{10}\bo_{00}}{-\delta}|\psi_0\ra 
	= \frac{\gamma}{\delta} \,\ba_{10}\prod_{\ell\ne0}\ba_{0\ell}|\text{vac}\ra,
\ee
and the relevant term of $|\psi_2\ra$ takes the form 
\bes&&
\bigg(\frac{\gamma\ba_{11}\bo_{01} \,J_0\ba_{01}\bo_{00}}{(2U_{01}+\delta)U_{00}}
-\frac{J_1\ba_{11}\bo_{10}\,\gamma\ba_{10}\bo_{00}}{(2U_{01}+\delta)\delta}
\bigg)|\psi_0\ra 
\nonumber\\&&
=\Big(\frac{2J_0}{U_{00}}-\frac{J_1}{\delta}\Big)\frac{\gamma}{2U_{12}+\delta}\,
\ba_{11}\prod_{\ell\ne0}\ba_{0\ell}|\text{vac}\ra.
\ees
With that we arrive at
\be
\la\ba_{10}\ba_{11}\ra \simeq 
\Big(\frac{2J_0}{U_{00}}-\frac{J_1}{\delta}\Big)\frac{2\gamma^2}{(2U_{12}+\delta)\delta}
\ee
leading to a sign change when both terms in the round bracket cancel each other. The change from 
positive to negative sign corresponds to the transition from MI$_0$ to MI$_{01}$ that is thus expected 
to occur at
 \be
\delta=\frac{U_{00}J_1}{2J_0}.
\ee

\section{Appendix B: Transition between MI$_{01}$ and MI$_{1}$}
We assume sharp filling $n=n_0+n_1=1$ and treat the tunnel terms as a perturbation. 
The unperturbed on-site problem is then given by 
\bes
\Ho_0 &=& \delta \delta \no_1 -\gamma (\ba_1\bo_0+\ba_0\bo_1) + 2U_{01}\no_0\no_1
\nonumber\\&&
	+\,\frac{U_{00}}{2}\no_0(\no_0-1)+\frac{U_{11}}{2}\no_1(\no_1-1)
\ees
where we dropped the site index $\ell$. In the subspace of one particle on a site the unperturbed 
ground state reads
\be
|\psi^{(0)}\ra= (a_0\ba_0+a_1\ba_1) |\text{vac}\ra,
\ee
with energy $\varepsilon_0=\frac{\delta}{2}-\frac{1}{2}[\delta^2+4\gamma^2]^{1/2}$ per site and
$a_1/a_0 = -\varepsilon_0/\gamma$, with $a_0^2+a_1^2=1$, giving in leading order perturbation theory
\be\label{eq:occupations}
n_0\simeq a_0^2 \quad\text{and}\quad n_1\simeq a_1^2.
\ee

In the course of the perturbation calculation we also need 
defect states with one particle less (a hole) and one extra particle. The hole state is simply given by 
the vacuum
\be
|\psi^{(h)}\ra= |\text{vac}\ra,
\ee
with energy $\varepsilon_h=0$.
The subspace with two particles on a site contains three states. For simplicity, we neglect the 
hybridization coupling $\gamma$ and approximate the eigenstates with an additional particle by states 
with sharp occupations of the orbitals $\alpha$,
\bes
|\psi^{(p20)}\ra &=& \frac{1}{\sqrt{2}}(\ba_0)^2|\text{vac}\ra,
\\
|\psi^{(p11)}\ra &=& \ba_0\ba_1|\text{vac}\ra,
\\
|\psi^{(p02)}\ra &=& \frac{1}{\sqrt{2}}(\ba_1)^2|\text{vac}\ra,
\ees
with unperturbed energies $\varepsilon_{p20}=U_{00}$, $\varepsilon_{p11}=2U_{01}+\delta$, and
$\varepsilon_{p02}=2U_{11}+2\delta$.

Re-introducing the site index $\ell$ the unperturbed ground state reads
\be
|\psi_0\ra =\prod_\ell  |\psi^{(0)}_\ell\ra = \prod_\ell (a_0\ba_{0\ell}+a_1\ba_{1\ell})|\text{vac}\ra.
\ee
The correlation function between the neighboring sites $0$ and $1$ obtains a finite value in the first 
order of the perturbation expansion with respect to tunneling
\bes\label{eq:0corr}
\la\psi|\ba_{\alpha 0}\bo_{\alpha 1}|\psi\ra 
&\simeq& \la\psi_0|\ba_{\alpha0}\bo_{\alpha1}|\psi_1\ra+\la\psi_1|\ba_{\alpha 0}\bo_{\alpha1}|\psi_0\ra
\nonumber\\
&=&2\la\psi_0|\ba_{\alpha 0}\bo_{\alpha1}|\psi_1\ra.
\ees
Here the relevant terms of the first-order state correction $|\psi_1\ra$ possess an extra 
particle in one of the three possible states on site 1 and a hole on site 0. These terms are related to 
the perturbation $-J_0\ba_{01}\bo_{00}+J_1\ba_{11}\bo_{10}$ and read
\bes
&&\bigg[
\frac{a_0^2J_0}{U_{00}-2\varepsilon_0}(\ba_{01})^2 
- \frac{a_1^2J_1}{U_{11}+2\delta-2\varepsilon_0}(\ba_{11})^2 
\nonumber\\&&
-\,\frac{a_0a_1(J_1-J_0)}{U_{01}+\delta-2\varepsilon_0}\ba_{11}\ba_{01} 
\bigg]
\prod_{\ell\ne0,1}(a_0\bo_0+a_1\bo_1)|\text{vac}\ra.
\ees
With this expression we obtain from Eqs.~(\ref{eq:0corr}) and (\ref{eq:occupations}) that
\bes\label{eq:leg0corr}
\la\ba_{00}\bo_{01}\ra 
	&\simeq& \frac{2n_0}{U_{00}(2U_{01}+\delta-2\varepsilon_0)}
		\Big[2n_0 J_0 (2U_{01}+\delta-2\varepsilon_0)
	\nonumber\\&&
		-\,n_1(J_1-J_0)(U_{00}-2\varepsilon_0)\Big]
\ees
and
\bes
\la\ba_{10}\bo_{11}\ra 
	&\simeq& -\frac{2n_1}{(U_{11}+2\delta-2\varepsilon_0)(2U_{01}+\delta-2\varepsilon_0)}
	\nonumber\\&&\times
		 \Big[2n_1 J_1 (2U_{01}+\delta-2\varepsilon_0)
	\nonumber\\&&\quad
		+\,n_0(J_1-J_0)(U_{11}+2\delta-2\varepsilon_0)\Big].
\ees
The transition from MI$_{01}$ to MI$_1$ is related to $\la\ba_{00}\bo_{01}\ra$ becoming negative. 
Approximating $2U_{01}+\delta-2\varepsilon_0\approx 2U_{01}$, which is consistent with our previous 
approximation to neglect $\gamma$ on doubly occupied sites, the transition is expected to occur when
\be
\frac{n_1}{n_0}\approx\frac{4J_0U_{01}}{(J_1-J_0)U_{00}}
\ee
or, equivalently, when
\be
n_0-n_1\approx\frac{(J_1-J_0)U_{00}-4J_0U_{01}}{(J_1-J_0)U_{00}+4J_0U_{01}}.
\ee

\bibliography{mybib}

\end{document}